\documentclass[a4paper,12pt]{article}
\usepackage{bm}
\usepackage{latexsym}
\usepackage{dcolumn}
\usepackage{lipsum, amsmath,amsfonts,amssymb}
\usepackage{graphicx,epsfig}
\usepackage{color}
\usepackage[multiple]{footmisc}

\def\l{\left}
\def\r{\right}
\def\DM{\mathrm{d}}

\def\eq#1{{Eq.~(\ref{#1})}}

\newcommand{\LL}{Lanczos-Lovelock }

\newcommand{\bh}{black hole }

\def \k{\overset{(0)}{  {K}^{\alpha}_{\beta} }}
\def \kp{\overset{(1)}{  {K}^{\alpha}_{\beta} }}

\usepackage{bm}
\usepackage{latexsym}
\usepackage{dcolumn}
\usepackage{amsmath,amsfonts,amssymb}
\usepackage{graphicx,epsfig}
\usepackage{color}
\def\eq#1{{Eq.~(\ref{#1})}}

\def\beq{\begin{equation}}
\def\eeq{\end{equation}}
\def\bea{\begin{eqnarray}}
\def\eea{\end{eqnarray}}
\def\benu{\begin{enumerate}}
\def\eenu{\end{enumerate}}
\def\nn{\nonumber}

\def\l{\left}
\def\r{\right}

\reversemarginpar

\def\DM{\mathrm{d}}

\begin{document}
\title{Membrane Paradigm and Horizon Thermodynamics in \LL Gravity}
\author{Sanved Kolekar$^{\ a,}$\footnote{sanved@iucaa.ernet.in}~ and Dawood Kothawala$^{\ b,}$\footnote{dawood.ak@gmail.com}\\ \\ \\
$^{a }$\ IUCAA, Pune University Campus, Ganeshkhind,\\
Pune 411007, India. \\
$^{b }$ \ Department of Mathematics and Statistics,\\
University of New Brunswick,\\
Fredericton, NB, Canada E3B 5A3
}

\date{\today}
\maketitle
\begin{abstract}

We study the membrane paradigm for horizons in Lanczos-Lovelock models of gravity in arbitrary $D$ dimensions and find compact expressions for the pressure $p$ and viscosity coefficients $\eta$ and $\zeta$ of the membrane fluid. We show that the membrane pressure is intimately connected with Noether charge entropy $S_{_{\mathrm{Wald}}}$ of the horizon when we consider a specific $m$-th order \LL model, through the relation $p^{(m)} A/T = \l[ (D-2m)/(D-2) \r] S^{(m)}_{_{\mathrm{Wald}}} $, where $T$ is the temperature and $A$ is the area of the horizon. Similarly, the viscosity coefficients are expressible in terms of entropy and quasi-local energy associated with the horizons. The bulk and shear viscosity coefficients are found to obey the relation $\zeta = -2 (D-3)/(D-2) \eta$.

\end{abstract}
\section{Introduction} \label{sec:intro}
The event horizon of a black hole is a one way membrane, albeit one whose existence is almost (i.e., apart from tidal forces which are small near the horizon for massive black holes) imperceptible to observers who fall freely across it. However, the horizon surface serves as a boundary of accessible region to observers who remain static outside the horizon. The so called membrane paradigm of black holes \cite{membranekipthorne, damour} essentially takes the viewpoint that, as far as such static observers are concerned, the black hole horizon can be replaced by a \textit{stretched} horizon, a membrane, endowed with specific physical properties which encode the presence of the inaccessible black hole region. Although originally developed to facilitate the easy comprehension of physics in black hole backgrounds for astrophysical applications, work over the last decade or so has indicated that physical properties of the membrane such as viscosity etc might have a deeper relevance, specifically from the point of view of holographic dualities which map gravitational systems to non-gravitational systems in one lower dimension. This has led to a renewed interest in the study of membrane paradigm.
\footnote{There is, however, an important point which, although well known, is worth repeating. In membrane paradigm, bulk viscosity of the membrane fluid turns out to be negative. Although why this happens is not a mystery (it can be traced to the teleological property of event horizons), it does betray the fact that a completely physical interpretation in terms of hydrodynamics is perhaps meaningless in a strict sense. On the other hand, works based on AdS-CFT yield zero bulk viscosity, which is a more pleasant aspect. The difference can actually be traced to the presence of the scale associated with hydrodynamic gradients, in addition to the length scale associated with surface gravity of the horizon. We thank the referee for pointing this out.}

More recently \cite{paddynavier, membranepressure} the membrane paradigm in Einstein gravity was revisited from a view point of the emergent gravity paradigm \cite{paddyaspects}. It was shown that the Damour-Navier Stokes equation governing the dynamics of the \bh membrane could be obtained starting from an action which could be given a thermodynamic interpretation as an entropy production rate when expressed in terms of thermodynamic variables such as temperature, entropy, pressure, etc of the horizon. Such an approach highlighted the unexplored deeper connection of the membrane paradigm with the horizon thermodynamics. However, a more formal approach would be to establish a correspondence between the thermodynamic variables and the quantities describing the membrane. Indeed one such relation was pointed out in \cite{membranepressure} connecting the membrane pressure $p_s$ to the entropy $S$ of the horizon through an equation of state as $p_sA = ST$ where $T$ is the Hawking temperature and $A$ is the area of the horizon. Further, it was shown that the membrane pressure $p_s$ also has a direct connection with the entropy density $s_{\rm shell}$ of a self-gravitating system of densely packed shells on the verge of forming a black hole. Here, one could question about the robustness of these relations for horizons in general. Hence the resolution would be to check whether these hold even for higher curvature theories of gravity such as \LL theories of gravity. Also, when one is working in Einstein gravity ( or Gauss-Bonnet gravity) the geometric structure of the fluid variables such as transport co-efficients of the membrane is not evident. For example, in Einstein's gravity, it is known that the shear viscosity $\eta$ is equal to $1/(16\pi)$, however the origin of this factor is not well understood in terms of the geometric properties of the horizon such as its relation with the intrinsic curvature of the horizon. To trace their geometric origins one needs to consider a higher curvature theory of gravity such as \LL theories of gravity of an arbitrary order $m$ and then proceed in a more formal general way.

For resolving these issues one first needs to establish the membrane paradigm for horizons for higher curvature theories of gravity, which according to our literature survey is still missing. In this note, we generalize the membrane paradigm for black holes to a particular class of higher curvature theories namely the \LL models of gravity. We follow the standard route of evaluating the surface stress tensor of the perturbed horizon surface, and then comparing it with the structure of the stress tensor for a viscous fluid \cite{maulik, membraneGB} to read off the pressure, transport co-efficients, etc. The geometric setup and the procedure is discussed in detail in section 2. As emphasized before, our main motivation in analyzing such higher curvature actions as \LL is to clearly highlight those characteristics of the fluid transport coefficients which are generic for arbitrary \LL actions, and separate them from the aspects which are peculiar to Einstein-Hilbert action. In the process we indicate a formal connection between thermodynamic properties of a black hole horizon, such as its Noether charge entropy and quasi-local energy, and the transport coefficients of the membrane fluid (see equations in the next paragraph). We also point out a very direct connection \cite{membranepressure} of the membrane pressure $p_s$ with the Wald entropy of the horizon and also the entropy density $s_{\rm shell}$ of a self-gravitating system of densely packed shells on the verge of forming a black hole, which has been recently worked out for \LL models in \cite{shellLL}. We briefly mention the summary of our results below.
\\
\\
\noindent SUMMARY OF THE RESULTS:
\\
\\
{\it A. Membrane transport coefficients}: 
\\
\\
We shall be mainly interested in membranes which have isotropic tangential stresses (pressure), which is only possible when {\it all directions}, {\it everywhere} within the horizon surface, are equivalent (that is, the horizon surface is maximally symmetric). Therefore, the cross-section of the horizon of the background geometry has curvature $^{(D-2)} \overset{(0)}{R}_{ABCD} = (\mathcal{K}/r^2) \,(\gamma_{AC}\gamma_{BD}-\gamma_{AD}\gamma_{BC})$, where $\mathcal{K}=\pm 1$ and $r$ is a constant of dimension $length$ which is related to the intrinsic Ricci scalar of the horizon cross-section as $\mathcal{K} (D-3)(D-2) r^{-2} =~ ^{(D-2)}\overset{(0)}{R}$. For simplicity of notation, we will assume $\mathcal{K}=+1$ in the intermediate steps, and only restore it in the final expressions (which is easy to do). Doing so will come in handy while discussing the case of planar horizons, for which we will simply put $\mathcal{K}=0$. 

With this assumption, we shall show that the horizon membrane has the stress tensor (see below for notations etc)
\begin{eqnarray}
t_{\alpha \beta}=\rho_s u_\alpha u_\beta + e^{(A)}_\alpha e^{(B)}_\beta \left( p_s
\gamma_{AB} - 2 \eta_s {\sigma_s}_{AB} - \zeta_s \theta_s \gamma_{AB}
\right)
\end{eqnarray}
with $p_s = \sum_m p^{(m)}_s$ etc., and
\begin{eqnarray}
\textrm {Pressure : } ~~p^{(m)}_s &=& \l( \frac{D_{2 m}}{D_2} \r) \l( \frac{\kappa}{2 \pi} \r) ~ 4 \pi m \alpha_m L^{D-2}_{m-1} \nonumber \\
\textrm {Energy density : }~ \rho^{(m)}_s &=& - \l( \frac{\theta_s}{\kappa}\r) p^{(m)}_s \nonumber \\
\textrm {Shear Viscosity : }~ \eta^{(m)}_s &=& \l( \frac{m \alpha_m r_H^2}{\mathcal{K} D_2 D_3} \r) \Biggl[ L^{D-2}_m - \l( \frac{2 \kappa}{r_H} \r) 
 (m-1) D_{2m}  L^{D-2}_{m-1} 
\Biggl] \nonumber \\
\textrm {Bulk Viscosity : }~ \zeta^{(m)}_s &=& - \l( \frac{2 D_3}{D_2} \r) \eta^{(m)}_s
\label{eq:memtranscoeff}
\end{eqnarray}
where we have introduced the notation $D_k \equiv (D-k)$ to avoid clutter, and $\alpha_m$ is the coupling constant of the $m^{\rm th}$ order \LL lagrangian
\begin{eqnarray}
L^D_m = \frac{1}{16 \pi} \frac{1}{2^m} \delta^{a_1 b_1 \ldots a_m b_m}_{c_1 d_1 \ldots c_m d_m} R^{c_1 d_1}_{a_1 b_1} \ldots R^{c_m d_m}_{a_m b_m}
\end{eqnarray}
\\
\\
{\it 2. Connection with thermodynamic properties of the horizon}: 
\\
\\
The transport coefficients above can be connected with the thermodynamic properties of the horizon in \LL gravity, by noting that the Wald entropy \cite{Wald} and quasi-local energy of the horizon in these theories are given by \cite{LLeom}
\begin{eqnarray}
S^{(m)}_{_{\mathrm{Wald}}} &=& 4 \pi m \alpha_m \int_H d\Sigma \; L^{D-2}_{m-1}
\nn \\
E^{(m)}_{\mathrm{Wald}} &=& \alpha_m \int^{\lambda} d \lambda \int_H \DM \Sigma \; L^{D-2}_{m}
\label{waldentropy}
\end{eqnarray}
This immediately implies that
\begin{eqnarray}
\frac{p_s}{T_{\rm{\infty}}} \;\; \underset{\delta \rightarrow 0}{\equiv} \;\; \sum \limits_{m=0}^{[D-1]/2} \l( \frac{D-2m}{D-2} \r) S^{(m)}_{\rm{Wald}} 
\end{eqnarray}
Similarly, it is obvious from the dependence on $L^{D-2}_{m-1}$ and $L^{D-2}_{m}$ factors appearing in the third and fourth equalities in Eqs.~(\ref{eq:memtranscoeff}) for the shear and bulk viscosities, that both $\eta$ and $\zeta$ are also expressible solely in terms of $S^{(m)}$ and $E^{(m)}$.

The paper is organized as follows. In section 2, we setup the notations and discuss the geometric setup required. In section 3, we generalize the membrane paradigm for a \LL theory of arbitrary order $m$ and obtain the expressions for pressure, transport co-efficients, etc. The results and some of their implications are discussed more fully in the final section 4. We shall work in $D$ spacetime dimensions. Latin indices $a, b, \ldots = 0$ to $(D-1)$, Greek indices $\mu, \nu, \ldots = 0, 2, 3, \cdots (D-1)$, and capitalized Latin indices $A, B, \ldots = 2$ to $(D-1)$.

\section{The setup} \label{sec:setup}

In this section we will briefly elaborate on the geometric set-up and the procedure to construct the membrane paradigm. 

We shall work in a $D$ dimensional spacetime containing an event horizon $H$ generated by the null geodesics $\bm l$. The surface gravity $\kappa$ is then determined through $\nabla_{\bm l} \bm l = \kappa \bm l$. We shall deal with a stationary background spacetime, in which case $\bm l$ corresponds to the timelike Killing vector $\bm \xi$ restricted to $H$ where $\xi^2=0$. For asymptotically flat spacetimes, the norm of $\bm l$ can be fixed at infinity and there is no ambiguity in the definition of $\kappa$. The stretched horizon $\mathcal{H}_s$ is then defined as a timelike surface infinitesimally close to $H$, and separated from $H$ along spacelike geodesics $\bm n$. This surface is spanned by a unit timelike vector field $\bm u$ and $(D-2)$ vectors $\bm e_{(A)}$'s. The correspondence between points on $H$ and $\mathcal{H}_s$ is made via ingoing null geodesics $\bm k$ normalized to have unit Killing energy, $\bm k \bm \cdot \bm l=-1$. The horizon $H$ is then located at $\lambda=0$, where $\lambda$ is the affine parameter along $\bm k$, related to the norm of $\bm \xi$ by $\lambda = - (1/2 \kappa) \xi^2$. We shall define $\delta = \sqrt{-\xi^2}$ to simplify notation and also to facilitate comparison with known results in the literature. In the limit $\lambda \to 0$, $\mathcal{H}_s \rightarrow H$ and $\bm u \to \delta^{-1} \bm l$ and $\bm n \to \delta^{-1} \bm l$.

The induced metric $h_{\mu \nu}$ on $\mathcal{H}_s$ is given by $h_{\mu \nu} = h_{ab} e^a_{(\mu)} e^b_{(\nu)}$, where $h_{ab} = g_{ab} - n_a n_b$ and $\bm e_{(\mu)}$'s are basis vectors spanning $\mathcal{H}_s$ ($\bm e_{(\mu)} \bm \cdot \bm n=0$). Similarly, the induced metric $\gamma_{AB}$ on the $(D\!-\!2)$-dimensional space-like cross-section of $\mathcal{H}_s$ orthogonal to $\bm u$ is given by $\gamma_{AB} = \gamma_{ab} e^a_{(A)} e^b_{(B)}$ where $\gamma_{ab} = h_{ab} + u_a u_b$ and $\bm e_{(A)}$'s are basis vectors spanning $\gamma$ ($\bm e_{(A)} \bm \cdot \bm u=0=\bm e_{(A)} \bm \cdot \bm n$). The extrinsic curvature of $\mathcal{H}_s$ is defined as $ K_{\mu \nu} = e^a_{(\mu)} e^b_{(\nu)} \nabla_a n_b$ and it is easy to verify that in the limit  $\delta \to 0$, we have: 
\begin{eqnarray}
K_{{\hat 0} {\hat 0}} &=& K_{\mu \nu} u^{\mu} u^{\nu} =- \delta^{-1} \kappa
\nn \\
K_{{\hat 0} A} &=& K_{\mu \nu} u^{\mu} e^{\nu}_{(A)}=0
\nn \\
K_{A B} &=& K_{\mu \nu} e^{\mu}_{(A)} e^{\nu}_{(B)}
\nn \\ &=& \delta^{-1} k_{AB}
\label{nulllimit},
 \end{eqnarray}
where $k_{AB}$ is the extrinsic curvature of the $(D-2)$-dimensional space-like cross-section of the true horizon $H$. This can be decomposed with respect to its trace free part as 
\begin{eqnarray}
k_{AB} = \sigma_{AB} +  \frac{1}{(D-2)} \theta\, \gamma_{AB},
\label{ksplit}
\end{eqnarray}
where $\gamma^{AB} \sigma_{AB}=0$, and $\theta$ and $\sigma_{AB}$ are the expansion scalar and shear respectively of the $\bm l$ congruence generating the horizon. Of course, for a Killing horizon, both of these vanish as $\delta^2$ and therefore $K_{{\hat 0} {\hat 0}}$ remains the only divergent contribution to $K_{\mu \nu}$. However, since we wish to investigate dissipative properties associated with the horizon, we shall perturb the background away from staticity and then obtain the membrane stress tensor to first order in perturbations to determine the transport coefficients. The extrinsic curvature can be written, upto linear order in perturbations, as
\begin{eqnarray}
 K^{\alpha}_{\beta} = \k + \kp 
\end{eqnarray}
where $\k$ is the unperturbed extrinsic curvature. Since we assumed the background geometry to be static, we have
\begin{eqnarray}
\overset{(0)}{K_{{\hat 0} {\hat 0}}} &=&- \kappa / \delta, \; \; \;
\overset{(0)}{K_{{\hat 0} A}} = 0, \; \;
\overset{(0)}{K_{A B}} = \left( \delta/r \right) \delta_{AB}
\label{zerothorder}
\end{eqnarray}
To study the additional divergences brought in when the horizon is perturbed, it is convenient to choose perturbations which do not affect $\overset{(0)}{K_{{\hat 0} {\hat 0}}}$.  
\footnote{As we shall see, our final relations will only have divergence of $O(\delta^{-1})$, and perturbing $\kappa$ will not lead to any qualitative difference (this can be easily shown). Formally, since we assume the background $\kappa$ to be non-zero, the perturbation in $\kappa$ is not gauge invariant and our choice corresponds to a gauge wherein $\delta \kappa=0$ \cite{membranekipthorne}.}
Therefore, we have
\begin{eqnarray}
\overset{(1)}{K_{{\hat 0} {\hat 0}}} &=& 0, \; \; \;
\overset{(1)}{K_{{\hat 0} A}} = 0, \; \;
\overset{(1)}{K_{A B}} = (1/\delta) k_{AB}
\label{firstorder}
\end{eqnarray}
where we have not bothered to put a ``$(1)$" over $k_{AB}$ since its unperturbed part is strictly zero. The expansion and shear parameters of the perturbed horizon will now introduce additional divergences in the stress tensor via $K_{AB}$.

We are now ready to use all the above facts to analyze the membrane stress tensor in \LL theories of gravity. For further details about the membrane paradigm and aspects of the perturbation scheme, we refer the reader to earlier literature on the subject \cite{membranekipthorne, maulik}, and also to the introductory sections in the recent paper \cite{membraneGB} which focuses on Einstein-Gauss-Bonnet theories.
\section{\LL gravity}

The $m^{\mathrm{th}}$ order \LL lagrangian $L_m$ is given by completely anti-symmetrised product of $m$ curvature tensors 
\begin{eqnarray}
L_m^{(D)} = \frac{1}{16 \pi} \frac{1}{2^m} \delta^{a_1 b_1 \ldots a_m b_m}_{c_1 d_1 \ldots c_m d_m} R^{c_1 d_1}_{~ a_1 b_1} \cdots R^{c_m d_m}_{~ a_m b_m}
\end{eqnarray}
For $m=1$, $L_m$ reduces to $(16 \pi)^{-1} R$, which is the Einstein-Hilbert lagrangian. The surface stress tensor in \LL theory is given by \cite{surfacestressLL}
\begin{eqnarray}
8 \pi t^{\nu}_{(m) \mu} &=&  \frac{\alpha _{m} m!}{2^{m+1}} \, \sum_{s=0}^{m-1} \widetilde{C}_s \, \widetilde{\pi}_{(s) \mu}^{\nu} 
\nonumber \\
\widetilde{\pi}^{\nu}_{(s) \mu} &=& \delta _{\lbrack
\mu \mu_{1}\cdots \mu_{2m-1}]}^{[\nu \nu_{1}\cdots \nu_{2m-1}]}\,{R}_{\nu_{1}\nu_{2}}^{\mu_{1}\mu_{2}}\cdots {R}_{\nu_{2s-1}\nu_{2s}}^{\mu_{2s-1}\mu_{2s}}%
\,K_{\nu_{2s+1}}^{\mu_{2s+1}}\cdots K_{\nu_{2m-1}}^{\mu_{2m-1}} 
\label{junctionconditionLLOLea}
\end{eqnarray}%
where the coefficients $\widetilde{C}_s$ are given by
\begin{equation}
\widetilde{C}_s = \frac{4^{m-s}}{s!\left( 2m-2s-1\right) !!}
\label{coefficient}
\end{equation}
Here $R_{\mu \rho \sigma \nu}$ denotes the projection of full spacetime curvature tensor on the $(D-1)$ dimensional hypersurface with normal $\bm n$. It turns out to be convenient to express this stress tensor in terms of the intrinsic curvature $\hat R_{\mu \rho \sigma \nu}$ of the hypersurface (i.e., defined using the induced metric $h_{\mu \nu}$), through the Gauss-Codazzi relation 
\begin{eqnarray}
\hat R_{\mu \rho \sigma \nu} = R_{\mu \rho \sigma \nu} - K_{\mu \sigma} K_{\rho \nu} + K_{\mu \nu} K_{\rho \sigma}
\end{eqnarray}
We can then write the surface stress energy tensor as 
\begin{eqnarray}
8 \pi t^{\nu}_{(m) \mu} &=&  \frac{\alpha _{m} m!}{2^{m+1}} \, \sum_{s=0}^{m-1} C_s \, \pi
_{(s) \mu}^{\nu} 
\nonumber \\
\pi^{\nu}_{(s) \mu} &=& \delta _{\lbrack
\mu \mu_{1}\cdots \mu_{2m-1}]}^{[\nu \nu_{1}\cdots \nu_{2m-1}]}\,{\hat R}_{\nu_{1}\nu_{2}}^{\mu_{1}\mu_{2}}\cdots {\hat R}_{\nu_{2s-1}\nu_{2s}}^{\mu_{2s-1}\mu_{2s}}%
\,K_{\nu_{2s+1}}^{\mu_{2s+1}}\cdots K_{\nu_{2m-1}}^{\mu_{2m-1}} 
\label{junctionconditionLL}
\end{eqnarray}%
with 
\begin{equation}
C_s=\sum_{q = s}^{m-1} \frac{ (-2)^{q-s} 4^{m-q} \; \binom{q}{s} }{q!\left( 2m-2q-1\right) !!}
\label{coefficient}
\end{equation}
One can check that for $m=1$ and $m=2$, the above expression reduces to that of the surface stress tensor in Einstein \cite{mtw} and Gauss-Bonnet \cite{junctionGB} theories respectively. 

As in the case of the membrane paradigm for Einstein gravity, we will interpret the $t_{\alpha \beta}$ as due to a fictitious matter source residing on the stretched horizon $\mathcal{H}_s$. Using the assumptions and expressions given above, and in the Sections (\ref{sec:intro}) and (\ref{sec:setup}), we will show that the surface stress energy tensor can be written in a form analogous to that of a viscous fluid, namely,
\begin{eqnarray}
t_{\alpha \beta}=\rho_s u_\alpha u_\beta + e^{(A)}_\alpha e^{(B)}_\beta \left( p_s
\gamma_{AB} - 2 \eta_s {\sigma_s}_{AB} - \zeta_s \theta_s \gamma_{AB}
\label{viscousfluidtensor}
\right)
\end{eqnarray}
and hence read off the corresponding energy density $\rho_s$, pressure $p_s$, shear viscosity co-efficient $\eta$ and co-efficient bulk viscosity $\zeta_s$ in terms of the geometrical quantities describing the horizon.

Before proceeding to describe the calculations, it is instructive to note that the determinant tensor has the property  
\begin{equation}
\delta^{{\hat 0} \alpha_1 \alpha_2 \cdots \alpha_n}_{{\hat 0} \beta_1 \beta_2 \cdots \beta_n} = \delta^{A_1 A_2 \cdots A_n}_{B_1 B_2 \cdots B_n} \times \l( \delta^{\alpha_1}_{A_1} \delta^{B_1}_{\beta_1} \ldots \delta^{\alpha_n}_{An} \delta^{B_n}_{\beta_n} \r)
\label{determinanttensor}
\end{equation}
That is, the presence of $0$ in each row of the determinant tensor forces all the other indices to take the values $2,3, \cdots (D-1)$. Further, keeping in mind that the ${\pi _{(s)}}^{\nu}_{\mu}$ in \eq{junctionconditionLL} is a polynomial of odd degree $\l[2(m-s)-1\r]$ in $K_{\mu \nu}$, it is easy to see that, to first order in perturbations, the only divergences in $t_{\alpha \beta}$ are of $O(\delta^{-1})$, and they arise from $s=(m-1)$ and $s=(m-2)$ terms in the series. This follows (upon simple counting) from Eqs.~(\ref{zerothorder}), (\ref{firstorder}), (\ref{junctionconditionLL}) and (\ref{determinanttensor}). All the other terms in the series are of $O(\delta)$ and hence vanish in the limit $\delta \rightarrow 0$. We give below an outline of the contribution of divergent terms, stating only the relevant expressions and skipping the cumbersome intermediate steps which are algebraically straightforward. We have also defined $\l( Q^{(D-1)}_m \r)^{\alpha \mu \beta \nu} = (1/m) \; \partial L^{(D-1)}_{m}/\partial \hat R_{\alpha \mu \beta \nu}$.

\begin{enumerate}

\item[A.] \textit{Divergence due to the zeroth order term}:

Only the $s = (m-1)$ term will contribute to the zeroth perturbative order. The corresponding contribution can be found in our recent work \cite{shellLL}, and turns out to be
\begin{eqnarray}
\hspace{-1cm} \mathrm{For~} s = (m-1):  \hspace{0.3in} 
\l[  \pi_{(s) {\hat 0} {\hat 0}}  \r]_{\rm div} &=& 0 
\nn \\
\nn \\
\l[  \pi^A_{(s) B}  \r]_{\rm div}&=& \delta^{-1}  \times 2^m \; 16 \pi \left( Q^{(D-1)}_m \right)^{{\hat 0} A}_{{\hat 0} B} \; \kappa
\end{eqnarray}

\item[B.] \textit{Divergence due to the first order term}:

At first order in the perturbations, we get divergences from $s=(m-1)$ and $s=(m-2)$. The contributions from these can also be found in a straightforward way, and are given by
\begin{eqnarray}
\hspace{0.5in} \mathrm{For~} s = (m-1):  \hspace{0.3in} 
\l[  \pi_{(s) {\hat 0} {\hat 0}}  \r]_{\rm div} &=& \delta^{-1}  \times 2^m \; 16 \pi \left( Q^{(D-1)}_m \right)^{{\hat 0} C}_{{\hat 0} D} k^D_C
\nn \\
\nn \\
\l[  \pi^A_{(s) B}  \r]_{\rm div} &=& \delta^{-1}  \times 2^m \; 16 \pi \left( Q^{(D-1)}_m \right)^{A C}_{B D} k^D_C
\nn \\
\nn \\
\hspace{0.5in} \mathrm{For~} s = (m-2):  \hspace{0.3in}
\l[ \pi_{(s) {\hat 0} {\hat 0}}  \r]_{\rm div} &=& 0
\nn \\
\nn \\
\l[  \pi^A_{(s) B}  \r]_{\rm div} &=& 6 \, \delta^{-1} \times \frac{\kappa}{r} \times 
\delta_{{\hat 0}FBD \; \bullet \bullet \cdots \star \star}^{{\hat 0}FAC \; \bullet \bullet \cdots \star \star} \, 
\underbrace{  \hat{R}_{\bullet \bullet}^{\bullet \bullet} \cdots \hat{R}_{\star \star}^{\star \star}  }_{2(m-2) \rm{factors}}
\; k^D_C
\nn \\
\end{eqnarray}
where the symmetry factor of $6$ in the last equation comes because we need to choose, from $3$ factors of $K_{\alpha \beta}$, one factor of $\overset{(1)}{K}_{AB}$ ($3$ ways) and one factor of $K_{uu}$ from the remaining $2$ $K_{\alpha \beta}$'s ($2$ ways), which can be done in a total of $3\times 2=6$ ways.
\end{enumerate}
The above equations can be further simplified by using the following identities (see appendix \ref{identitiesproof} for proof)
\begin{eqnarray}
 \left( Q^{(D-1)}_m \right)^{0A}_{0B} &=& \l( \frac{D_{2m}}{2 D_2} \r) L^{D-2}_{m-1} \; \delta^A_B 
 \label{identity1}  \\
\left( Q^{(D-1)}_m \right)^{AC}_{BD} &=& \frac{1}{2} \l( \frac{L^{D-2}_{m}}{^{D-2}R} \r) \; \delta^{AC}_{BD} 
\label{identity2} \\
%
%
\delta_{
B{\hat 0}DF \; \bullet \bullet \cdots \star \star}^{A{\hat 0}CE \; \bullet \bullet \cdots \star \star } \, 
\underbrace{  \hat{R}_{\bullet \bullet}^{\bullet \bullet} \cdots \hat{R}_{\star \star}^{\star \star}  }_{2(m-2) \rm{factors}}
&=& \l( \frac{ 2^{m-2} D_{2m}}{D_4} \r) \left( \frac{16\pi L^{D-2}_{m-1}}{^{D-2}R} \right) \; \delta^{ACE}_{BDF}
\label{identity3}
\end{eqnarray}
where we remind the notation $D_k=(D-k)$ introduced earlier. We can now collect the $O(\delta^{-1})$ terms in the total stress tensor $t_{\alpha \beta} = \overset{(0)}{t}_{\alpha \beta} +\overset{(1)}{t}_{\alpha \beta}$ upto first order in perturbation to get
\begin{eqnarray}
 t^{{\hat 0}}_{{\hat 0}} &=& \delta^{-1} \times 2 m \alpha_m \; \l( \frac{D_{2m}}{D_2} \r) L^{D-2}_{m-1} \; \theta 
\\
t^{A}_{B} &=& \delta^{-1} \times 2 m \alpha_m 
\Biggl\{  
\l( \frac{D_{2m}}{D_2} \r) L^{D-2}_{m-1} \; \kappa \; \delta^A_B
+ \l( \frac{L^{D-2}_{m}}{^{D-2}R} \r) \; k^A_B
\nn \\ 
&&- 2 (m-1) D_{2m} \left( \frac{L^{D-2}_{m}}{^{D-2}R}\right) \frac{\kappa}{r_H} k^A_B
\Biggl\}
\end{eqnarray}
We have therefore accomplished our main task of finding the divergent contributions to surface stress tensor $t_{\mu \nu}$ of the membrane to linear order in horizon perturbations. The only divergence is of $O(\delta^{-1})$, which is precisely the same as in Einstein theory, and can therefore be regulated in the same way as is done in conventional membrane paradigm for Einstein gravity. That is, we simply multiply by $\delta$ and obtain the finite part of $t_{\mu \nu}$, with the regularization having the interpretation in terms of infinite redshift at the horizon. Now writing $k^A_B$ in terms of the traceless part $\sigma^A_B$ and trace $\theta$ as in \eq{ksplit}, we can easily express the stress tensor in the required form
\begin{eqnarray}
t_{\alpha \beta}=\rho^{(m)}_s u_\alpha u_\beta + e^{(A)}_\alpha e^{(B)}_\beta \left( p^{(m)}_s
\gamma_{AB} - 2 \eta^{(m)}_s {\sigma_s}_{AB} - \zeta^{(m)}_s \theta_s \gamma_{AB}
\right)
\end{eqnarray}
and read off the pressure, transport co-efficients, etc as
\begin{eqnarray}
\textrm {Pressure : } ~~p^{(m)}_s &=& \l( \frac{D_{2 m}}{D_2} \r) \l( \frac{\kappa}{2 \pi} \r) ~ 4 \pi m \alpha_m L^{D-2}_{m-1} \nonumber \\
\textrm {Energy density : }~ \rho^{(m)}_s &=& - \l( \frac{\theta_s}{\kappa}\r) p^{(m)}_s \nonumber \\
\textrm {Shear Viscosity : }~ \eta^{(m)}_s &=& \l( \frac{m \alpha_m }{^{^{(D-2)}}R} \r) \Biggl[ L^{D-2}_m - \l( \frac{2 \kappa}{r_H} \r) 
 (m-1) D_{2m}  L^{D-2}_{m-1} 
\Biggl] \nonumber \\
\textrm {Bulk Viscosity : }~ \zeta^{(m)}_s &=& - \l( \frac{2 D_3}{D_2} \r) \eta^{(m)}_s
\label{finalexpressioneta}
\end{eqnarray}
where we again remind of our notation $D_k \equiv (D-k)$. One can check that for Einstein's gravity $m=1$ and Gauss-Bonnet gravity $m=2$, the above expressions reduce to the corresponding  expressions in the literature \cite{membranekipthorne, membraneGB}. 

\subsection{The $\eta/s$ ratio:}

Using the expression of Wald entropy given in \eq{waldentropy}, we can determine the shear viscosity to entropy density ratio to be
\begin{eqnarray}
 \frac{\eta^{(m)}}{s^{(m)}} &=& \l( \frac{1}{4\pi \; ^{^{D-2}}R } \r) \Biggl[ \frac{L^{D-2}_m}{L^{D-2}_{m-1}} - \l( \frac{2 \kappa}{r_H} \r) 
 (m-1) D_{2m} \Biggl]
 \nn \\
 \nn \\
 &=& \l( \frac{D_{2m}}{4\pi D_2 D_3} \r) \Biggl[ D_{2m+1} - 2 \left( \kappa r_H \right) 
 (m-1) \mathcal{K}^{-1} \Biggl]
 \nn \\
\nn \\ 
&:=& \frac{c_m}{4 \pi} \; \rm{(say)}
\end{eqnarray}
where we have used the relation $L^{D-2}_{m} = \l( \mathcal{K}/r^2 \r) D_{2m} D_{2m+1} L^{D-2}_{m-1}$ which holds for a  $(D-2)$ dimensional maximally symmetric spacetime. We can therefore write
\begin{eqnarray}
 \frac{\eta}{s} &=& \frac{1}{4 \pi} \l[ 
 \frac{1+\sum \limits_{m=2}^{m_c} c_m {\bar s}_m}{1+\sum \limits_{m=2}^{m_c} {\bar s}_m \textcolor{white}{c_m} }
\r]
\end{eqnarray}
where $m_c=[(D-1)/2]$ and ${\bar s}_m := s_m/s_1=4 s_m$. One can now investigate the ratio $\eta/s$ on a case-by-case basis, considering specific solutions with horizons in \LL theory. Specifically, for {\it planar} horizons, $\mathcal{K}=0$, and we note that in this case, $c_m \rightarrow (\ldots)\mathcal{K}^{-1}$. 
\footnote{A couple of clarifying points while dealing with planar horizons are in order: (i) although ($\kappa r_H$) would in general depend on $\mathcal K$, we expect it to be finite for $\mathcal K=0$, and (ii) the most general form of a planar horizon metric, consistent with our assumption of staticity, is: $\DM s_H^2 = \Omega^2(r/L_i) \l( \DM x^2 + \DM y^2 + \ldots \r)$, where $\Omega(r/L_i)$ is a dimensionless function constructed out of ratios of the coordinate $r$ and any other length scale(s) $L_i$ appearing in the metric. For known solutions which are asymptotically AdS, $\Omega(r/L_i) \equiv r/\ell$, $\ell$ being the AdS scale.}
Therefore, since ${\bar s}_m \rightarrow (\ldots)\mathcal{K}^{m-1}$, only the $m=2$ term in the sum in the numerator survives for $\mathcal{K}=0$. One is therefore lead to the conclusion that no Lovelock term other than Gauss-Bonnet contributes to the $\eta/s$ ratio for the planar ($\mathcal{K}=0$) case; this contribution is given by
\begin{eqnarray}
\hspace{-1cm} \mathrm{For~\mathcal{K}=0} :  \hspace{0.5in} 
 \frac{\eta}{s} &=& \frac{1 + [c_2 {\bar s}_2]_{\mathcal{K}=0}}{4 \pi} 
\nn \\
&=& \frac{1}{4 \pi} \l[ 1 - 4 \alpha_2 (D-4) \l( \frac{\kappa}{r_H} \r) \r]
\end{eqnarray}
However, although only the Gauss-Bonnet contributes for the planar case, it must be noted that $\eta/s$ will, in general, depend on other Lovelock coupling constants as well, through the ratio $\l(\kappa/r_H\r)$ which must be calculated using a specific $\mathcal{K}=0$ solution of the full Lovelock action. Similar result has been noted in the literature \cite{shu, edelstein} in the context of AdS-CFT correspondence; the difference being the boundary used which is the AdS infinity whereas we have considered the boundary to be the stretched horizon. However, it has been argued in \cite{rgflow} that the $\eta/s$ ratio for the two cases are related by a (RG) flow equation in Enstein's gravity and it would indeed be interesting to check whether such a flow exist in \LL theory, since our result for the membrane paradigm, in conjunction with the AdS-CFT results for \LL gravity, does indicate such a possible connection in the $\mathcal{K}=0$ case. Further, in the {\it special case} of a black brane solution in Einstein-Gauss-Bonnet theory, we have $\kappa/r_H=(D-1)/2 \ell^2$, in which case the result above matches exactly with the one obtained in the context of AdS-CFT (see \cite{membraneGB} and references within). Using the expression for $\eta/ s$, it would be interesting to check whether it obeys or violates the KSS bound \cite{KSSbound}, $\eta/ s \geq 1/(4\pi)$, by explicitly computing it for black hole solutions known for \LL gravity. For $\mathcal{K}=0$ case, the KSS bound is found to be violated in the AdS-CFT context in Einstein-Gauss Bonnet gravity \cite{membraneGB, kssboundGB, kssboundGBIP} and also in a general \LL theory of gravity \cite{shu, kssboundLL}.

\section{Conclusions and Discussion}

We have studied the membrane paradigm for black hole horizons in \LL gravity theories and found various transport coefficients associated with the membrane fluid, assuming staticity and maximal symmetry of the horizon surface. Beyond a generalization of known results to \LL actions, there are other important implications of the calculations presented here, which we briefly discuss in this concluding part. 

First, from our final results, the connection between various transport coefficients and the thermodynamic properties of the horizon, such as temperature, entropy and quasi-local energy, becomes immediately obvious. To the best of our knowledge, such a direct connection has not been discussed previously in the literature. Indeed, it would be very difficult to establish any such connection unambiguously if one is working within the context of Einstein theory itself (or, for that matter, any particular $m$-th \LL term), since the formal structure of the transport coefficients is not then apparent. It would be extremely interesting if such a connection holds for horizons in more general class of gravity theories.

The second aspect which the calculation presented here highlights is the connection between Wald entropy $S_{\rm{Wald}}$ of the horizon and the pressure $p_s$ of the horizon membrane. This relation is encoded in the form of an equation of state, 
\begin{eqnarray}
\frac{p_s A}{T_{\rm{loc}}} \;\; \underset{\delta \rightarrow 0}{\equiv} \;\; \sum \limits_{m=1}^{[D-1]/2} \l( \frac{D-2m}{D-2} \r) S^{(m)}_{_{\rm{Wald}}}
\end{eqnarray}

Finally, the algebraic steps leading to the above result are precisely the same as those leading to the evaluation of entropy of a self-gravitating configuration of densely packed shells, held in equilibrium with its own ``acceleration" radiation, and on the verge of becoming a black hole. The entropy $S_{\rm{matter}}$ in this case was evaluated by Oppenheim \cite{areascalingoppenheim} in the context of Einstein theory, motivated by an ``operational" approach first discussed by Pretorius et. al. \cite{areascalingisrael}. In a recent publication \cite{shellLL}, the calculation of $S_{\rm{matter}}$ was done for \LL models of gravity, and it was shown that several new features arise when one goes beyond Einstein theory -- for example, while $S_{\rm{matter}}$ correctly gives the Bekenstein-Hawking entropy in Einstein theory, it is {\it not} equal to Wald entropy (which is a generalization of Bekenstein-Hawking entropy to higher derivative gravity theories) for the \LL theories. In particular, $S_{\rm{matter}}$ near the horizon turns out to be $PA/T_{\rm{loc}}$, and in fact $P \equiv p_s$ appearing in the membrane paradigm. This connects up the operational approach to horizon entropy discussed in \cite{shellLL} with the membrane paradigm for the black holes. In fact, it was while generalizing the former approach in \cite{shellLL} that the connection with membrane paradigm became apparent, thereby providing the motivation for the present work.

Before we conclude, we must also mention that the relation between bulk and shear viscosities for \LL order $m$, $\zeta^{(m)}/\eta^{(m)} = -2 (D-3)/(D-2)$,  is {\it independent} of $m$, which immediately implies $\zeta/\eta = -2 (D-3)/(D-2)$ for an action which is a sum of certain \LL terms. This corroborates the results already known for Einstein gravity, and recently also established for Einstein-Gauss-Bonnet case \cite{membraneGB}. It would be worth investigating whether there is a deeper reason for the robustness of this ratio. Also, from the final expression for $\eta$ in \eq{finalexpressioneta}, it is clear that the shear viscosity (and thereby the ratio $\eta/s$ which has been a point of focus in many recent investigations) depends not only on horizon entropy density $s \propto L^{D-2}_{m-1}$, but also on quasi-local energy $E \propto L^{D-2}_m$ of the horizon. In fact, for the Einstein case $m=1$, the entropy contribution to $\eta$ vanishes, and the result $\eta/s=(4 \pi)^{-1}$ is only a consequence of the quasi-local energy of the horizon! Once again, this brings into sharp focus the relevance of going beyond Einstein's theory to gain a better understanding of several aspects of horizon thermodynamics and membrane paradigm in Einstein theory itself.

\section*{Acknowledgements}
We thank Olivera Miskovic and Rodrigo Olea for useful clarifications regarding the surface stress tensor in \LL gravity. We thank T. Padmanabhan for useful comments on the draft of this paper. We thank the referee for his/her comments on improving the clarity of notations. SK is supported by a Fellowship from the Council of Scientific and Industrial Research (CSIR), India. The research of DK is funded by NSERC of Canada, and Atlantic Association for Research in the Mathematical Sciences (AARMS).

\vspace{0.2cm}

\appendix

\section{Proof of identities in Eqs.~(\ref{identity1}) - (\ref{identity3}) } \label{identitiesproof}
To prove the required identities, we will use the following two relations concerning \LL actions of order $m$ in $D$ dimensions \cite{LLeom}
\begin{eqnarray}
{\mathcal G}^{i (D)}_{j(m)} &=&  - \frac{1}{2} \frac{1}{16 \pi} \frac{1}{2^m} \delta^{i a_1 b_1 \ldots a_m b_m}_{j c_1 d_1 \ldots c_m d_m} R^{c_1 d_1}_{~ a_1 b_1} \cdots R^{c_m d_m}_{~ a_m b_m}
\nn \\
{\mathcal G}^{i (D)}_{i(m)} &=& - \frac{D-2m}{2} L^D_m \label{eomproperty}
\end{eqnarray}
where ${\mathcal G}^{i (D)}_{j(m)}$ is the equation of motion tensor for \LL action; for e.g., for $m=1$, ${\mathcal G}^{i (D)}_{j(1)} = (16 \pi)^{-1} G^i_j$ where $G^i_j$ is the Einstein tensor. Note that for maximally symmetric spacetime, which is the case we are considering for the $(D-2)$ submanifold corresponding to the $(D-2)$ dimensional horizon, we have $G^2_2 = G^3_3 = \cdots = G^{D-1}_{D-1}$ due to isotropy. Hence we can write the second relation in \eq{eomproperty} for equation of motion tensor for the $(D-2)$ dimensional maximally symmetric subspace as 
\begin{eqnarray}
{\mathcal G}^{A (D-2)}_{B(m)} &=& - \frac{\delta^A_B}{2} \left[ \frac{(D-2)-2m}{D-2} \right] L^{D-2}_m \label{eommaxsymm}
\end{eqnarray}
Using the above identity it is easy to prove \eq{identity1} as
\begin{eqnarray}
\left( Q^{0B}_{0A} \right)^{(D-1)}_{m} &=& \frac{1}{16 \pi} \frac{1}{2^m} \delta _{\lbrack
0A \mu_{1}\cdots \mu_{2m-2}]}^{[0B \nu_{1}\cdots \nu_{2m-2}]}\,\hat{R}_{\nu_{1}\nu_{2}}^{\mu_{1}\mu_{2}}\cdots \hat{R}_{\nu_{2m-3}\nu_{2m-2}}^{\mu_{2m-3}\mu_{2m-2}} \nonumber \\
&=& \frac{1}{16 \pi} \frac{1}{2^m} \delta _{\lbrack
A A_{1}\cdots A_{2m-2}]}^{[B B_{1}\cdots B_{2m-2}]}\,\hat{\hat{R}}_{B_{1}B_{2}}^{A_{1}A_{2}}\cdots \hat{\hat{R}}_{B_{2m-3}B_{2m-2}}^{A_{2m-3}A_{2m-2}} \nonumber \\
&=& - {\mathcal G}^{B (D-2)}_{A(m-1)} = \frac{1}{2} \left( \frac{D-2m}{D-2} \right) L^{D-2}_{m-1} \; \delta^A_B
\end{eqnarray}
where to obtain the second equality we have used \eq{determinanttensor} and replaced $\hat{R}^{A B}_{C D}$ with the intrinsic curvature $\hat{\hat{R}}^{A B}_{C D}$ defined completely in terms of the induced metric $\gamma_{AB}$ of the horizon. Such a replacement is valid since the corresponding extrinsic curvature vanishes due to our assumption of staticity. The last equality then follows from using \eq{eommaxsymm}.
To prove the second identity of \eq{identity2} note that
\begin{eqnarray}
\left( Q^{BD}_{AC} \right)^{(D-1)}_{m} &=& \frac{1}{16 \pi} \frac{1}{2^m} \delta _{\lbrack
AC \mu_{1}\cdots \mu_{2m-2}]}^{[BD \nu_{1}\cdots \nu_{2m-2}]}\,\hat{R}_{\nu_{1}\nu_{2}}^{\mu_{1}\mu_{2}}\cdots \hat{R}_{\nu_{2m-3}\nu_{2m-2}}^{\mu_{2m-3}\mu_{2m-2}} \nonumber \\
&=& \frac{1}{16 \pi} \frac{1}{2^m} \delta _{\lbrack
AC A_{1}\cdots A_{2m-2}]}^{[BD B_{1}\cdots B_{2m-2}]}\,\hat{\hat{R}}_{B_{1}B_{2}}^{A_{1}A_{2}}\cdots \hat{\hat{R}}_{B_{2m-3}B_{2m-2}}^{A_{2m-3}A_{2m-2}} \nonumber \\
&=& \left( Q^{BD}_{AC} \right)^{(D-2)}_{m} = M^{(D-2)}_{m} \delta^{BD}_{AC}
\end{eqnarray}
where to obtain the second equality we have used $\hat{R}^{\hat{{\hat 0}} \mu}_{\rho \nu} = 0$ which is true due to (i) static nature of the background geometry and (ii) maximal symmetry of the $(D-2)$ dimensional subspace and  we have again replaced $\hat{R}^{A B}_{C D}$ with the intrinsic curvature $\hat{\hat{R}}^{A B}_{C D}$. Further, in the last equality we have again exploited maximal symmetry to get a $\delta^{BD}_{AC}$ proportionality with a proportionality constant $M^{(D-2)}_{m} $ which we can determine by contracting the above equation with $\hat{\hat{R}}^{A B}_{C D}$ to get
\begin{eqnarray}
L^{(D-2)}_{m} = \left( Q^{BD}_{AC} \right)^{(D-2)}_{m} \hat{\hat{R}}^{A C}_{B D} &=&  M^{(D-2)}_{m} \delta^{BD}_{AC} \hat{\hat{R}}^{A C}_{B D} = 2 M^{(D-2)}_{m} \; ^{D-2}R
\end{eqnarray}
This completes the proof of identity in \eq{identity2}. To prove the third identity of \eq{identity3} we again use the same procedure and write 
\begin{eqnarray}
 \delta _{\lbrack
A0CE \mu_{1}\cdots \mu_{2m-4}]}^{[B0DF \nu_{1}\cdots \nu_{2m-4}]}\,\hat{R}_{\nu_{1}\nu_{2}}^{\mu_{1}\mu_{2}}\cdots \hat{R}_{\nu_{2m-5}\nu_{2m-4}}^{\mu_{2m-5}\mu_{2m-4}} &=&  \delta _{\lbrack
ACE A_{1}\cdots A_{2m-4}]}^{[BDF B_{1}\cdots B_{2m-4}]}\,\hat{\hat{R}}_{B_{1}B_{2}}^{A_{1}A_{2}}\cdots \hat{\hat{R}}_{B_{2m-5}B_{2m-4}}^{A_{2m-5}A_{2m-4}} \nonumber \\
&=& N^{(D-2)}_{m-1} \delta^{BDF}_{ACE}
\end{eqnarray}
with the proportionality constant $N^{(D-2)}_{m-1} $ determined by contracting both sides of the above expression by  $\hat{\hat{R}}^{A B}_{C D}$ and then taking a trace to get
\begin{eqnarray}
(16\pi)2^{m-1} \frac{D-2m}{D-2} L^{D-2}_{m-1} = N^{(D-2)}_{m-1} (16\pi)2^{1} \frac{D-4}{D-2} L^{D-2}_{1} 
\end{eqnarray}
This completes the proof of identity in \eq{identity3}.




\begin{thebibliography}{25}

\bibitem{membranekipthorne}
Thorne K S et al., \textit{Black Holes: The Membrane Paradigm} (Yale University Press, 1986)

\bibitem{damour} 
T. Damour (1979), \textit{Quelques propri\'et\'es m\'ecaniques, \'electromagn\'etiques,
thermo\-dy\-na\-mi\-ques et quantiques des trous noirs},
Th\`ese de doctorat d'\'Etat, Universit\'e Paris 6. (available at http://www.ihes.fr/~damour/Articles/);
T. Damour (1982), \textit{Surface effects in black hole physics}, \textit{Proceedings of the Second Marcel Grossmann Meeting on General Relativity}, Ed. R. Ruffini, North Holland , p. 587.

\bibitem{paddynavier}
T.~Padmanabhan (2011), \textit{Phys. Rev.},\textbf{ D83}, 044048 [arXiv:1012.0119] 

\bibitem{membranepressure}
Sanved Kolekar and T. Padmanabhan (2011), \textit{Action principle for the Fluid-Gravity correspondence and emergent gravity}, [arXiv:1109.5353]

\bibitem{paddyaspects}
T.~Padmanabhan (2010), \textit{Rept. Prog. Phys.}, \textbf{73}, 046901 [arXiv:0911.5004];
T. Padmanabhan (2011), \textit{ J. Phys. Conf. Ser.}, \textbf{306},  012001 [arXiv:1012.4476] 

\bibitem{maulik} 
Maulik Parikh and Frank Wilczek (1998), \textit{Phys. Rev.}, \textbf{D 58}, 064011 [arXiv:gr-qc/9712077]

\bibitem{membraneGB} 
Ted Jacobson, Arif Mohd, Sudipta Sarkar (2011), \textit{The Membrane Paradigm for Gauss-Bonnet gravity}  [arXiv:1107.1260] 

\bibitem{Zaslavskii} 
J. Lemos and O. Zaslavskii (2011), \textit{Phys. Rev.} \textbf{D 84}, 064017 [arXiv:1108.1801]

\bibitem{shellLL}
Sanved Kolekar, Dawood Kothawala and T. Padmanabhan (2011), \textit{Two Aspects of Black hole entropy in Lanczos-Lovelock models of gravity} [arXiv:1111.0973]


\bibitem{Wald}
Robert Wald (1993), \textit{Phys. Rev.}, \textbf{D48}, 3427-3431 [arXiv:gr-qc/9307038]; Vivek Iyer and Robert Wald (1994), \textit{Phys. Rev. D}, \textbf{50}, 846-864 [arXiv:gr-qc/9403028] 


\bibitem{LLeom}
Dawood Kothawala and T. Padmanabhan (2009), \textit{Phys. Rev.}, \textbf{D79}, 104020 [arXiv:0904.0215]


\bibitem{surfacestressLL}
Olivera Miskovic and Rodrigo Olea (2007), \textit{JHEP}, \textbf{10}, 028 [arXiv:0706.4460]; Claudio Teitelboim and Jorge Zanelli (1987), \textit{Class. Quantum Grav.}, \textbf{4}, L125-L129

\bibitem{mtw}
Misner, Thorne, Wheeler, \textit{Gravitation}, W. H. Freeman and Company (1973).

\bibitem{junctionGB}
Stephen C. Davis (2003), \textit{Phys.Rev.}, \textbf{D67}, 024030 [arXiv:hep-th/0208205]; Nathalie Deruelle and Cristiano Germani (2003), \textit{Nuovo Cim.}, \textbf{B118}, 977-988 [arXiv:gr-qc/0306116]; Claude Barrabes and W. Israel (2005), \textit{Phys.Rev.}, \textbf{D71}, 064008 [arXiv:gr-qc/0502108]

\bibitem{shu}
Fu-Wen Shu (2010), \textit{Phys.Lett.}, \textbf{B685}, 325-328 [arXiv:0910.0607]

\bibitem{edelstein}
Xian Camanho and Jose Edelstein (2010), \textit{Causality in AdS/CFT and Lovelock theory} [arXiv:0912.1944]

\bibitem{rgflow}
Irene Bredberg, Cynthia Keeler, Vyacheslav Lysov and Andrew Strominger (2011), \textit{JHEP}, \textbf{1103}, 141 [arXiv:1006.1902]

\bibitem{KSSbound}
P. Kovtun, D. T. Son and A. O. Starinets (2003), \textit{JHEP}, \textbf{0310}, 064 [arXiv:hep-th/0309213]; P. Kovtun, D. T. Son and A. O. Starinets (2005), \textit{Phys. Rev. Lett.}, \textbf{94}, 111601 [arXiv:hep-th/0405231].

\bibitem{kssboundGB}
Mauro Brigante, Hong Liu, Robert Myers, Stephen Shenker and Sho Yaida (2008), \textit{Phys.Rev.}, \textbf{D77}, 126006 [arXiv:0712.0805]

\bibitem{kssboundGBIP}
I. P. Neupane (2009), \textit{Int.J.Mod.Phys.} \textbf{A24}, 3584 [arXiv:0904.4805]; I. P. Neupane and N. Dadhich (2009), \textit{Class.Quant.Grav.}, \textbf{26}, 015013  [arXiv:0809.1818]

\bibitem{kssboundLL}
Xian Camanho, Jose Edelstein and Miguel Paulos (2011), \textit{JHEP}, \textbf{1105}, 127 [arXiv:1010.1682] 



\bibitem{areascalingoppenheim}
Jonathan Oppenheim (2001), \textit{Phys. Rev.}, \textbf{D 65}, 024020 [arXiv:0105101] 



\bibitem{areascalingisrael}
F. Pretorius, D. Vollick and W. Israel (1998), \textit{Phys. Rev.}, \textbf{D 57}, 6311 [arXiv:9712085] 



\end{thebibliography}
\end{document}